\documentclass[prr,reprint,twocolumn,superscriptaddress]{revtex4-1}
\usepackage{graphicx,amssymb,amsmath,bbold,bm,xcolor}
\usepackage{hyperref}
\graphicspath{{img/}}
\usepackage{mathtools}
\usepackage[caption=false]{subfig}

\def\arucl{$\alpha$-RuCl$_3$}

\begin{document}
	
\title{Emergence of nematic paramagnet via quantum order-by-disorder \\
and pseudo-Goldstone modes in Kitaev magnets}
\author{Matthias Gohlke}
\affiliation{Theory of Quantum Matter Unit, Okinawa Institute of Science and Technology Graduate University,
  Onna-son, Okinawa 904-0495, Japan}
\author{Li Ern Chern}
\affiliation{Department of Physics, University of Toronto, Toronto,
  Ontario M5S 1A7, Canada}
\author{Hae-Young Kee}
\affiliation{Department of Physics, University of Toronto, Toronto,
  Ontario M5S 1A7, Canada}
\affiliation{Canadian Institute for Advanced Research/Quantum
  Materials Program, Toronto, Ontario MSG 1Z8, Canada}
\author{Yong Baek Kim}
\affiliation{Department of Physics, University of Toronto, Toronto,
  Ontario M5S 1A7, Canada}
\affiliation{School of Physics, Korea Institute for Advanced Study,
  Seoul 130-722, Korea}
\date{\today}     
	
\begin{abstract}

The appearance of nontrivial phases in Kitaev materials exposed to an external magnetic field
has recently been a subject of intensive studies.
Here, we elucidate the relation between the field-induced ground states of the
classical and quantum spin models proposed for such materials,
by using the infinite density matrix renormalization group (iDMRG)
and the linear spin wave theory (LSWT).
We consider the $K \Gamma \Gamma'$ model, where $\Gamma$ and $\Gamma'$ are off-diagonal
spin exchanges on top of the dominant Kitaev interaction $K$.
Focusing on the magnetic field along the $[111]$ direction,
we explain the origin of the nematic paramagnet,
which breaks the lattice-rotational symmetry and exists in an extended window of magnetic field,
in the quantum model.
This phenomenon can be understood as the effect of quantum order-by-disorder
in the frustrated ferromagnet with a continuous manifold of degenerate ground states
discovered in the corresponding classical model.
We compute the dynamical spin structure factors using a matrix operator based time evolution
and compare them with the predictions from LSWT.
We, thus, provide predictions for future inelastic neutron scattering experiments on Kitaev
materials in an external magnetic field along the $[111]$ direction.
In particular, the nematic paramagnet exhibits a characteristic pseudo-Goldstone mode
which results from the lifting of a continuous degeneracy via quantum fluctuations.

\end{abstract}
	\maketitle

%%%%%%%%%%%%%%%%%%%%%%%%%%%%%%%%%%%%%%%%%%%%%%%%%%%%%%%%%%%%%%%%%%%%%%%%%%%%%%%%%%%%%%%%%%
%%% Introduction
%%%%%%%%%%%%%%%%%%%%%%%%%%%%%%%%%%%%%%%%%%%%%%%%%%%%%%%%%%%%%%%%%%%%%%%%%%%%%%%%%%%%%%%%%%
\section{Introduction} \label{scn:intro}

Recently, there has been a surge of interest in Kitaev materials \cite{rau_spinorbit_2016,trebst_kitaev_2017,winter_models_2017,hermanns_physics_2018,takagi_kitaev_2019,motome_hunting_2019}
due to the promise for the discovery of a Kitaev spin liquid \cite{kitaev_anyons_2006}.
The presence of strong spin-orbit coupling in these materials gives rise to bond-dependent
interactions between effective spin-$1/2$ moments on the underlying honeycomb lattice \cite{jackeli_mott_2009,witczak-krempa_correlated_2014,nussinov_compass_2015}.
In addition to a substantial Kitaev interaction, there exist other types of spin-exchange interactions \cite{rau_generic_2014} in these materials, which ultimately stabilize a magnetic order instead of the desired quantum spin liquid.
A paradigmatic example of Kitaev materials is $\alpha$-RuCl$_3$ \cite{plumb_aRuCl3_2014},
which has a zigzag (ZZ) ordered ground state \cite{sears_magnetic_2015,johnson_monoclinic_2015,banerjee_proximate_2016,sears_ferromagnetic_2019}.
Upon applying a magnetic field, however, the ZZ order vanishes
\cite{johnson_monoclinic_2015,ponomaryov_unconventional_2017,wolter_fieldinduced_2017,wang_magnetic_2017,banerjee_excitations_2018,hentrich_large_2018,lampenkelley_fieldinduced_2018},
while a half-quantized thermal Hall conductivity is observed \cite{kasahara_majorana_2018}.
This would be consistent with the theoretical prediction of the non-abelian chiral spin liquid
in the pure Kitaev model under a magnetic field \cite{kitaev_anyons_2006}.
It leads to the speculation that the field-induced phase is indeed the quantum spin liquid.

The experiment mentioned above \cite{kasahara_majorana_2018}
and recent inelastic neutron scattering experiments \cite{banerjee_proximate_2016,banerjee_neutron_2017,winter_probing_2018,banerjee_excitations_2018,balz_finite_2019}
have motivated a number of numerical studies \cite{gordon_theory_2019,jiang_fieldinduced_2019,kaib_kitaev_2019,chern_magnetic_2020,lee_magneticfield_2019}
in an effort to identify possible nontrivial phases in theoretical spin models proposed for realistic materials.
For instance, the $K \Gamma \Gamma'$ model, which is considered as a minimal model describing a broad class of Kitaev materials including $\alpha$-RuCl$_3$,
has been studied with an external magnetic field.
Exact diagonalization on a $24$-site cluster shows that the Kitaev spin liquid (KSL) is stable in a window of fields slightly tilted from the $[111]$ direction \cite{gordon_theory_2019}.
In a field along the $[111]$ axis, however, a tensor network study reveals that KSL is confined to the vicinity of the pure Kitaev model,
while novel nematic paramagnetic states, which breaks the lattice-rotational symmetry,
occupy a significant portion of the phase diagram at intermediate fields \cite{lee_magneticfield_2019}.
On the other hand, the classical model exhibits a plethora of magnetic orders with large unit cells,
and a ferromagnetic phase frustrated by the $\Gamma$ interaction and the field,
before the system becomes fully polarized \cite{janssen_magnetization_2017,janssen_heisenberg_2019,chern_magnetic_2020}.

\begin{figure*}
    \includegraphics[width=\linewidth]{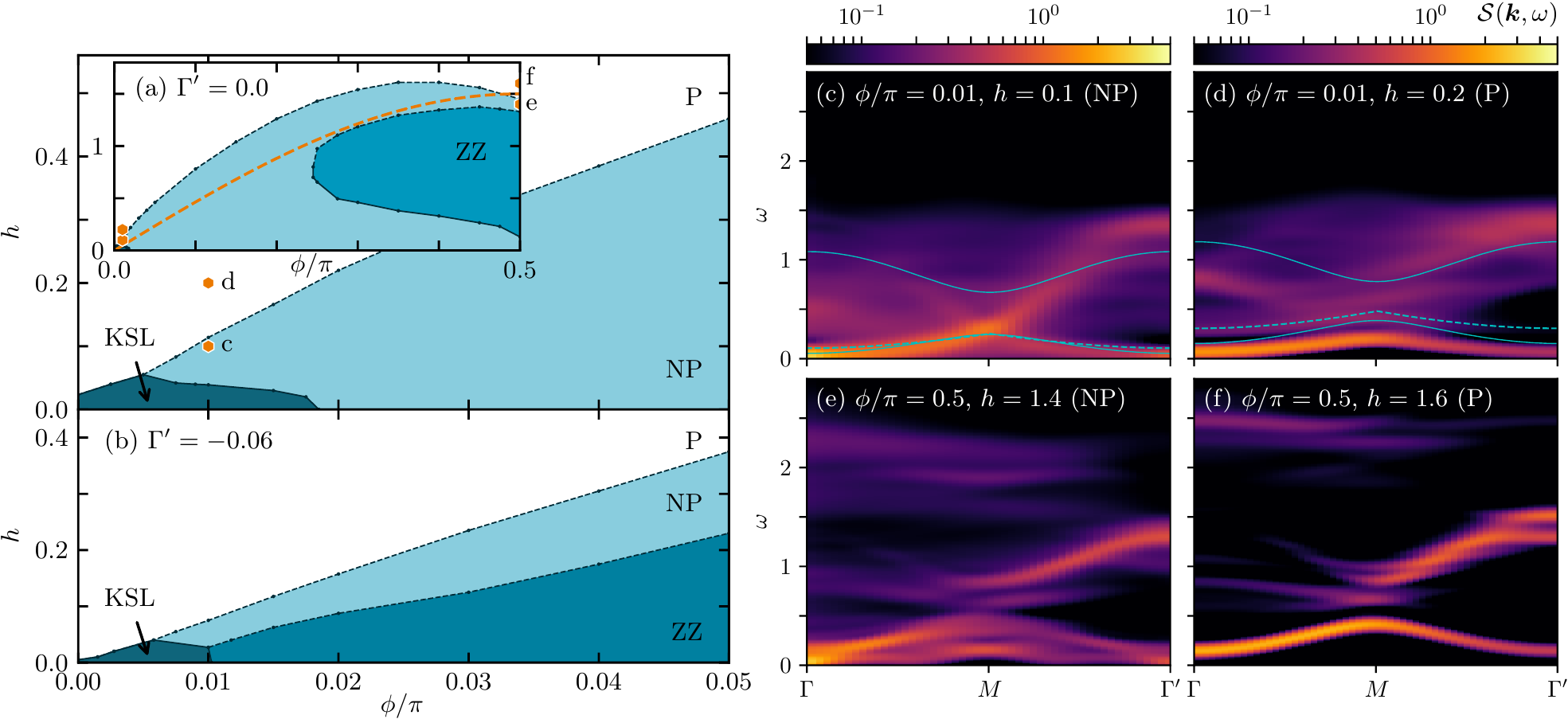}
    \caption{
        Phase diagram of the $K \Gamma \Gamma'$ model in a magnetic field along the $[111]$ axis
        when (a) $\Gamma'=0$ and (b) $\Gamma'=-0.06$.
        In both cases, the Kitaev spin liquid (KSL) is observed
        only in a small region near $\phi \rightarrow 0$ and $h \rightarrow 0$.
        A large fraction of the phase diagram is occupied by a nematic paramagnet (NP) that 
        spontaneously breaks the $C_3$ lattice-rotation symmetry.
        The inset illustrates the extension of NP down to zero field
        and right to the pure $\Gamma$ model.
        A long-range zigzag (ZZ) ordered phase appears at $\Gamma/|K| \gtrsim 1$
        at intermediate fields.
        The orange dashed line indicates the transition between the fully-polarized phase and
        frustrated ferromagnet based on ansatz \eqref{eqn:O3spin} in the corresponding classical model.
        A finite $\Gamma'<0$, i.e. (b) $\Gamma'=-0.06$, induces ZZ at low fields
        such that NP occurs at intermediate fields between ZZ and the longitudinally-polarized paramagnet (P) that is adiabatically connected to the fully-polarized limit.
        The phase diagram (a,b) is obtained on the \emph{rhombic-2} cylinder geometry
        with $L_\mathrm{circ}=10$ (see Fig.~\ref{fig:geom}).
        (c-f) Dynamical spin structure factors $\mathcal S(k,\omega)$ of P and NP near the upper critical field at two opposite limits of the $K \Gamma$ model as illustrated by the orange hexagons in (a).
        In both limits, the magnon band at low energies, which is apparent in P, becomes diffuse in NP, whereas the continuum at higher energies persists across the transition P to NP.
        The characteristic feature of NP is the increased spectral weight at $\Gamma$ and at low energies, which is indicative of the pseudo-Goldstone mode.
        Remarkably, the continuum exhibits a very different structure in both limits. 
        }
    \label{fig:pd}
\end{figure*}
In this work, we investigate the origin of the nematic paramagnet in the 
quantum $K \Gamma \Gamma'$ model and its relation to the classical ground states,
employing the infinite density matrix renormalization group (iDMRG) approach
\cite{white_density_1992,mcculloch_infinite_2008,phien_infinite_2012}
and linear spin wave theory (LSWT).
Our main findings are:
\begin{enumerate}
    \item The nematic paramagnetic state arises due to quantum order-by-disorder effect
          in the continuous manifold of degenerate classical ground states,
          dubbed frustrated ferromagnet, in the corresponding classical model. 
    \item The dynamical structure factors for the nematic paramagnetic state
          reveal the existence of a pseudo-Goldstone mode
          resulting from the quantum order-by-disorder effect.
          Our predictions for the dynamical structure factors can be compared to
          future neutron scattering experiments to confirm the nematic paramagnetic state.
          The dynamical structure factors in the corresponding regions in the phase diagram
          are shown in Fig.~\ref{fig:pd}.
\end{enumerate}

Furthermore, in the process of achieving these goals,
we attempt to resolve the differences and establish meaningful connections between
the results of different numerical studies,
with the understanding that different numerical methods possess their own limitations
and offer complementary information.
We elaborate all of these findings in a concise fashion in the rest of the introduction.
Further and complete details can be found in the main text.

We first investigate the phase diagram of the $K\Gamma\Gamma'$ model in a magnetic field along the $[111]$ direction.
In agreement with Ref.~[\onlinecite{lee_magneticfield_2019}], we find the KSL to be confined to a small corner of the phase diagram,
while a nematic paramagnet (NP) appears at intermediate fields before the system enters the longitudinally-polarized paramagnet (P) at high field
that is adiabatically connected to the fully polarized limit.
The NP phase breaks lattice-rotational symmetry but preserves translational symmetry.
In that context, it is worth emphasizing that NP is not the type of \emph{quantum spin-nematic} associated with the breaking of spin-rotation symmetry
\cite{blume_biquadratic_1969,andreev_spin_1984,chubukov_chiral_1991,shannon_nematic_2006}.
We expect that NP is more likely to be observed than the KSL due to the much wider extent of the former in parameter space. 

Next, we reveal the origin of NP by demonstrating that it arises from 
a quantum order-by-disorder effect \cite{shender_antiferromagnetic_1982,henley_ordering_1989} 
in the frustrated ferromagnet (FF) found in the classical model \cite{chern_magnetic_2020}.
In the classical limit, the FF appears in a window of intermediate magnetic fields between 
a series of magnetically ordered phases and the polarized state at low and high fields respectively.
The spin orientation in the FF possesses an azimuthal symmetry, which results in a $U(1)$
degenerate manifold of states.
Upon the inclusion of zero point quantum fluctuations via the linear spin wave theory (LSWT) \cite{rau_pseudogoldstone_2018},
a discrete set of azimuthal angles is selected, lifting the $U(1)$ degeneracy.
The selection of azimuthal angles, which depends on the field strength,
is consistent with our iDMRG results as well as the magnetizations of the NP states identifed by the tensor network study \cite{lee_magneticfield_2019}.
Furthermore, the series of large unit cell orders \cite{chern_magnetic_2020} observed 
at low fields in the classical model is replaced by NP in the quantum model.

We then present the dynamical spin structure factors (DSF) of the NP and P phases of the quantum model, using the matrix product operator based time evolution (tMPO) \cite{zaletel_timeevolving_2015},
and compare them to the spin wave dispersions as obtained in the semiclassical LSWT approach.
In P at high fields, the DSF clearly shows two magnon bands.
The magnon excitation gap shrinks as the field decreases towards the transition into NP.
At the critical field, the magnetic excitations become gapless at the $\Gamma$ and $\Gamma'$ points
in the reciprocal space, which can be interpreted as the onset of the pseudo-Goldstone mode
associated with the lifted $U(1)$ degeneracy \cite{rau_pseudogoldstone_2018} upon entering NP.
In NP, while there exists relatively well defined magnon excitations at lower energies,
a broad continuum forms at higher energies, which is reminiscent of that as seen in inelastic 
neutron scattering experiments
\cite{banerjee_proximate_2016,banerjee_neutron_2017,winter_probing_2018,banerjee_excitations_2018,balz_finite_2019}.
Approaching the KSL, the excitation continuum becomes much broader, which may be considered
as a signature of the proximate KSL \cite{banerjee_proximate_2016,yamaji_clues_2016,gohlke_dynamics_2017}.
We suggest to perform neutron scattering experiments with the $[111]$ magnetic field to test
our predictions for the pseudo-Goldstone mode and the nematic paramagnetic state. 

%%%%%%%%%%%%%%%%%%%%%%%%%%%%%%%%%%%%%%%%%%%%%%%%%%%%%%%%%%%%%%%%%%%%%%%%%%%%%%%%%%%%%%%%%%
%%% Model
%%%%%%%%%%%%%%%%%%%%%%%%%%%%%%%%%%%%%%%%%%%%%%%%%%%%%%%%%%%%%%%%%%%%%%%%%%%%%%%%%%%%%%%%%%
\section{Model} \label{scn:mod}
We study spin-$\frac{1}{2}$ degrees of freedom on a honeycomb lattice with bond-dependent interactions. The Hamiltonian of interest is given by
\begin{align}
	\mathcal H = &~~ \sum_{\langle i,j \rangle_\gamma} K_\gamma S^\gamma_i S^\gamma_j 
		 + \sum_{\mathclap{\langle i,j \rangle_\gamma, \alpha,\beta \neq \gamma}} 
\Gamma_\gamma \left( S^\alpha_i S^\beta_j + S^\beta_i S^\alpha_j\right) \nonumber \\
		&+ \sum_{\mathclap{\langle i,j \rangle_\gamma, \alpha,\beta \neq \gamma}} 
\Gamma'_\gamma \left( S^\alpha_i S^\gamma_j + S^\beta_i S^\gamma_j + S^\gamma_i 
S^\alpha_j + S^\gamma_i S^\beta_j \right) \nonumber \\
		&- h \sum_i \hat{\mathbf{h}} \cdot  \mathbf{S}_i~, 
        \label{eqn:Ham}
\end{align}
where $\langle i,j\rangle_\gamma$ are neighboring sites
connected by a bond with label $\gamma \in \{x,y,z\}$. The first term is the 
Kitaev or bond-dependent Ising exchange, which, in the absence of other interactions, 
stabilize a quantum spin liquid \cite{kitaev_anyons_2006}. The 
second and third terms are the off-diagonal $\Gamma$ and $\Gamma'$ exchanges. 
Classically, the pure $\Gamma$ model is known to host a spin liquid 
\cite{rousochatzakis_classical_2017}, but its quantum ground state is still
under debate \cite{rousochatzakis_classical_2017,wang_one_2019a,luo_gapless_2019}.
With dominant $K<0$ and $\Gamma>0$ interactions, a finite $\Gamma'<0$ exchange stabilizes the
long-range ordered zigzag phase at zero field
\cite{rau_trigonal_2014a, rau_generic_2014}.
The $h$ term describes the Zeeman coupling of the 
spins to an external magnetic field along the direction $\hat{\mathbf{h}}$.
Here, we consider a trigonometric parametrization of the Kitaev and $\Gamma$
interactions such that 
\begin{equation}
    K= -\cos\phi \qquad \textrm{and} \qquad \Gamma = \sin \phi~.
\end{equation}
We focus on the range $0 \le \phi \le \pi/2$,
fix $\Gamma'$ to be either $0$ or $-0.06$
\footnote{The choice of $\Gamma'=-0.06$ is based on
    (a) $\Gamma'$ being sufficiently large to induce ZZ order and 
    (b) $\Gamma'$ being sufficiently small to keep a small region of KSL in the phase diagram when using iDMRG.
    Consequently, our choice for $\Gamma'$ differs slightly from the ones used in other works \cite{gordon_theory_2019,lee_magneticfield_2019,chern_magnetic_2020}.},
and apply the field in the $[111]$ direction.
The $[111]$ field retains the $C_3$ symmetry of the $K\Gamma\Gamma'$ model.

%%%%%%%%%%%%%%%%%%%%%%%%%%%%%%%%%%%%%%%%%%%%%%%%%%%%%%%%%%%%%%%%%%%%%%%%%%%%%%%%%%%%%%%%%%
%%% Classical U(1) degeneracy and Order-by-disorder
%%%%%%%%%%%%%%%%%%%%%%%%%%%%%%%%%%%%%%%%%%%%%%%%%%%%%%%%%%%%%%%%%%%%%%%%%%%%%%%%%%%%%%%%%%
\section{Classical $U(1)$ degeneracy and quantum order-by-disorder} 
\label{scn:semiclassicalanalysis}
\begin{figure*}
    \includegraphics[width=\linewidth]{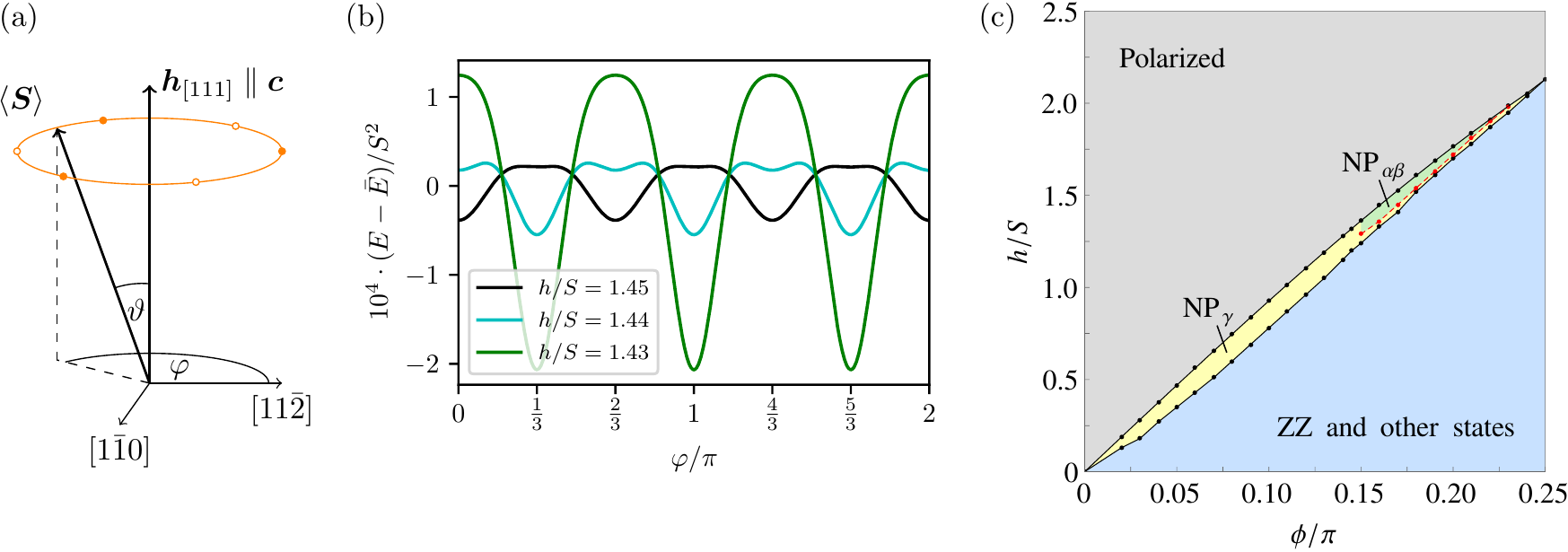}
    \caption{
        (a) Canting of the classical spin away from the $[111]$ field,
        its $U(1)$ manifold (orange line) of degenerate states,
        and the two sets of azimuthal angles selected by quantum fluctuations (empty and filled circles). 
        (b) Lifting of the classical $U(1)$ degeneracy via quantum order-by-disorder. 
        The azimuthal dependence of energy, Eq.~\eqref{eqn:qobd},
        at $\phi/\pi=0.17$, $\Gamma'=0$ and $h \in \{1.43,1.44,1.45\}$. 
        where the average energy $\bar E = 1/{2\pi} \int_{0}^{2\pi} d\varphi E(\varphi)$
        has been subtracted.
        As the field is tuned, there is a transition from one set of azimuthal angles to the other.
        (c) Frustrated ferromagnet (FF) in the phase diagram of the classical
        $K\Gamma$ model ($\Gamma'=0$) subjected to a magnetic field in $[111]$ direction. 
        FF is stabilized between the fully polarized state and other nontrivial ordering patterns,
        e.g. ZZ and 18-site order \cite{chern_magnetic_2020}.
        Upon incorporating quantum fluctuations via Eq.~\eqref{eqn:qobd}, FF turns into the nematic paramagnet (NP).
        The subscripts $\gamma$ and $\alpha\beta$ indicate that quantum fluctuations
        lift the classical $U(1)$ degeneracy differently.
    }
    \label{fig:u1_panel}
\end{figure*}

We start by considering the classical limit %($S\longrightarrow \infty$)
of \eqref{eqn:Ham} with $O(3)$ spins.
At high fields a narrow window with ferromagnetic order exists
where the spins are uniformly aligned but canted away from the $[111]$ field.
Such a canting originates from the competition 
between the field and the $\Gamma$ interaction
\cite{chern_magnetic_2020,janssen_magnetization_2017,janssen_heisenberg_2019},
thus motivating the name \emph{frustrated ferromagnet} (FF) for such a phase.

Assuming a ferromagnetic ansatz, i.e.~for all sites $i$, we write
\begin{equation} \label{eqn:O3spin}
    \mathbf{S}_i = S \left( \sin \vartheta \cos \varphi \; \hat{\mathbf{a}}
                 + \sin \vartheta \sin \varphi \; \hat{\mathbf{b}}
                 + \cos \vartheta \; \hat{\mathbf{c}} \right)
                 \equiv \mathbf{S} ,
\end{equation}
where $\hat{\mathbf{a}}$, $\hat{\mathbf{b}}$ and $\hat{\mathbf{c}}$ are 
the unit vectors along the $[11\bar{2}]$, $[\bar{1}10]$ and $[111]$ directions (or the 
$a$, $b$ and $c$-axes) respectively, $\vartheta = [0,\pi]$ is the polar angle measured from the 
$c$-axis, while $\varphi = [0, 2\pi)$ is the azimuthal angle in the $ab$-plane measured 
from the $a$-axis.

Substituting \eqref{eqn:O3spin} into \eqref{eqn:Ham},
the energy turns out to depend only on $\vartheta$ but not on $\varphi$.
The critical field, above which the spins are fully polarized, is calculated to be $h_\mathrm{crit} = 3 \left( \Gamma + 2\Gamma' \right) S$.
Below $h_\mathrm{crit}$, the spins are canted away from the $[111]$ direction by $\vartheta = \cos^{-1} \left( h/h_\mathrm{crit} \right)$,
yielding an emergent $U(1)$ manifold of degenerate ground states.
When choosing a particular value of $\varphi$ out of the $U(1)$ degenerate manifold, the spins spontaneously break a continuous symmetry.
As a consequence, the spin wave spectrum of the FF order exhibits a Nambu-Goldstone mode \cite{goldstone_broken_1962}.

We further investigate whether the classical $U(1)$ degeneracy is lifted 
by quantum fluctuations, demonstrating a concept known as quantum order-by-disorder \cite{shender_antiferromagnetic_1982,henley_ordering_1989}.
Within LSWT, the quantum correction to the energy is given by \cite{rau_pseudogoldstone_2018}
\begin{equation}
    E = S^2 \left[ \left( 1+\frac{1}{S} \right) E_\mathrm{cl} + \frac{1}{S} \frac{1}{2} 
\sum_{\mathbf{k}n} \omega_{\mathbf{k}n} + O \left( \frac{1}{S^2} \right) \right] ~,
    \label{eqn:qobd}
\end{equation}
where $E_\mathrm{cl}$ is the classical energy and $\omega_{\mathbf{k}n}$ is the magnon dispersion.

Our calculation shows that either of the following sets of azimuthal angles,
$\lbrace \pi/3, \pi, 5\pi/3 \rbrace$ and $\lbrace 0, 2\pi/3, 4\pi/3 \rbrace$,
are more energetically favorable than others (see Fig.~\ref{fig:u1_panel}(a) and (b)).
Within each set, the three angles yield the same energy,
i.e.~the $U(1)$ degeneracy is broken down to a $C_3$ degeneracy.
For $\varphi=\pi/3$, $\pi$ and $5\pi/3$, the spins cant towards the cubic 
axes $[010]$, $[001]$ and $[100]$ respectively,
leading to $S^\alpha = S^\beta < S^\gamma$, where $\alpha, \beta, \gamma \in \{x,y,z\}$
denote the spin components and are also related to the labels of bonds. 
As we will show later, the selected state is related to the NP phase occurring in the quantum model.
We therefore denote this set as NP$_\gamma$.
On the other hand, for $\varphi=0$, $2\pi/3$, and $4\pi/3$,
the spins cant towards the $[110]$, $[011]$ and $[101]$ directions respectively,
leading to $S^\alpha = S^\beta > S^\gamma$, which we denote as NP$_{\alpha\beta}$.
The magnetization of NP$_\gamma$ (NP$_{\alpha\beta}$) coincide with that of the NP1 (NP2) phase reported in Ref.~[\onlinecite{lee_magneticfield_2019}]. 

As the field is tuned, our quantum order-by-disorder computation reveals a transition from NP$_\gamma$ to NP$_{\alpha\beta}$, e.g.~Fig.~\ref{fig:u1_panel}(b).
Fixing $\Gamma'=0$, we indicate the location of the FF, as well as the corresponding NP states selected by quantum fluctuations,
in the classical phase diagram of \eqref{eqn:Ham} as shown in Fig.~\ref{fig:u1_panel}(c).

%%%%%%%%%%%%%%%%%%%%%%%%%%%%%%%%%%%%%%%%%%%%%%%%%%%%%%%%%%%%%%%%%%%%%%%%%%%%%%%%%%%%%%%%%%
%%% DMRG Ground state phase diagram
%%%%%%%%%%%%%%%%%%%%%%%%%%%%%%%%%%%%%%%%%%%%%%%%%%%%%%%%%%%%%%%%%%%%%%%%%%%%%%%%%%%%%%%%%%
\section{Quantum spin-$\frac{1}{2}$ ground state} \label{scn:pd} 
\begin{figure}
    \includegraphics[]{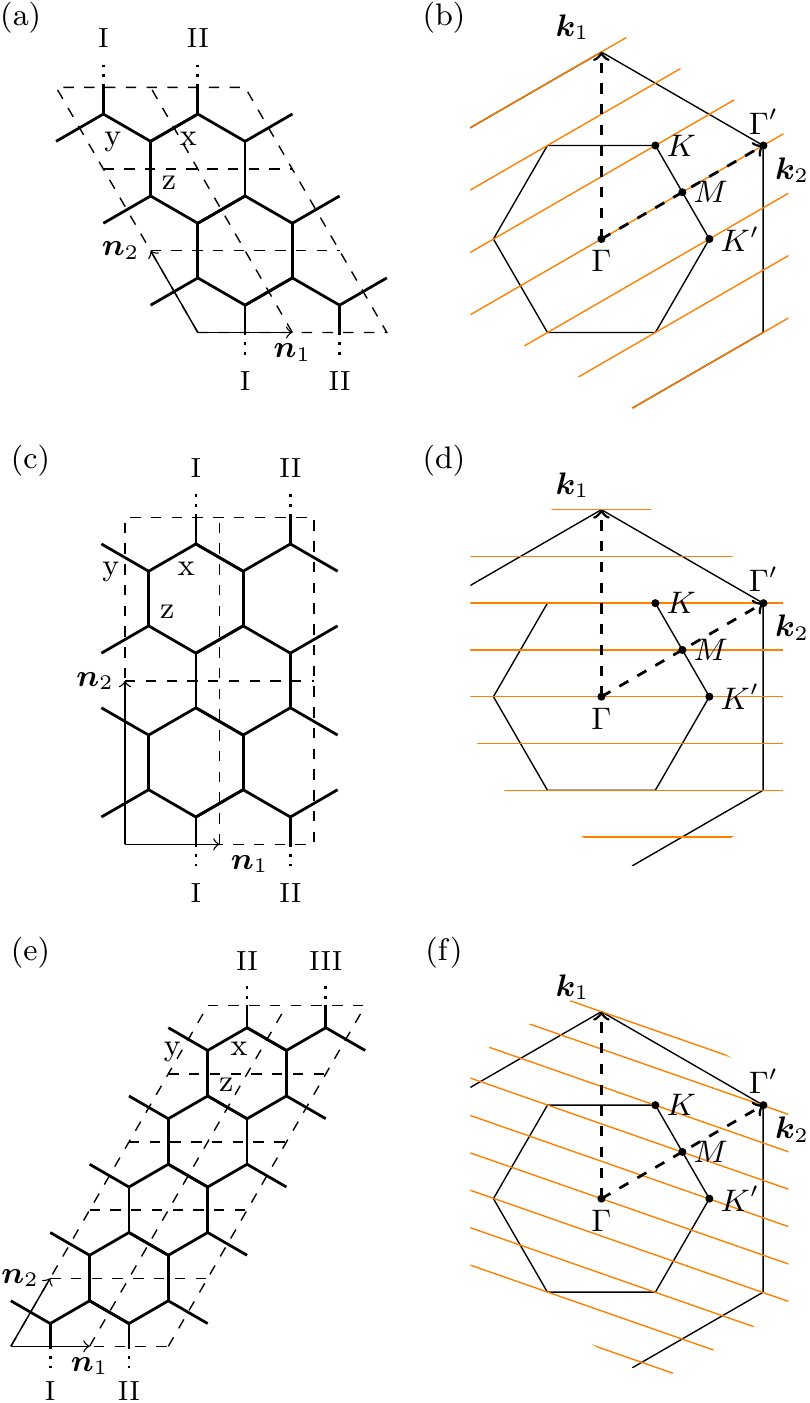}
    \caption{
        In order to enable the use of iDMRG to obtain the ground state wave function in the form of matrix product state (MPS),
        the two-dimensional honeycomb lattice is mapped to an infinitely-long cylinder.
        The following geometries are used:
        (a) \emph{rhombic} with a circumference of $L_\mathrm{circ}=6$ and $12$ (not shown) sites and (b) its allowed momenta in reciprocal space,
        (c,d) \emph{rectangular} with $L_\mathrm{circ}=8$ and $12$ (not shown),
        and (e,f) \emph{rhombic-2} with $L_\mathrm{circ}=10$.
        %The last geometry uses a shifted boundary condition
        %such that the matrix product operator representation of the Hamiltonian is
        %translationally invariant along a chain with a two-site unit cell.
        The geometries are chosen such that they 
        (1) capture the $K$-point in reciprocal space, at which gapless Majorana-Dirac cones of the Kitaev model with isotropic coupling strength occur \cite{kitaev_anyons_2006},
        and (2) the flux-decoration of the classical $\Gamma$ model \cite{rousochatzakis_classical_2017} is not frustrated.
        %Please see Appendix \ref{app:dmrggeometry} for more details.
        }
    \label{fig:geom}
\end{figure}
We consider now the quantum limit of \eqref{eqn:Ham} with spin-$\frac{1}{2}$ constituents. 
Fig.~\ref{fig:pd} shows the phase diagram near the KSL for (a) $\Gamma'=0$ and (b) $\Gamma'=-0.06$,
as well as (inset) the entire range $0 < \phi/\pi < 0.5$ for $\Gamma'=0$.
We employ iDMRG \cite{white_density_1992,mcculloch_infinite_2008,phien_infinite_2012}
on cylinders with various geometries and circumferences as illustrated in Fig.~\ref{fig:geom}.
The phase diagram is obtained on the \emph{rhombic-2} geometry with $L_\mathrm{circ}=10$ sites
and twisted boundary conditions.
The other geometries with upto $L_\mathrm{circ}=12$ are used checking against possible classical ordering patterns,
see Sec.~\ref{subscn:zerofield} for more details.

Firstly, we find the KSL to be confined to a 
small corner near the Kitaev limit, while a large fraction 
of the phase diagram is occupied by either NP 
that spontaneously breaks the lattice-rotational symmetry
or ZZ long-range magnetic order that further breaks translational symmetry.

At high fields, the ground state is a longitudinally-polarized paramagnet (P) 
that is adiabatically connected to the fully-polarized limit.
The system enters NP upon lowering the field
except at very small values of $\Gamma$ where a direct transition into the KSL occurs.
Similar to the FF phase, NP is characterized by a canting of the field-induced magnetic moment away from the $[111]$ field.
In contrast to FF, NP extends down to $h=0$ if $\Gamma'=0$, where the magnetic moment vanishes.
Since at zero field NP has already broken the $C_3$ symmetry,
as indicated by anisotropic bond energies,
a field-induced magnetic moment is no longer protected to align in the $[111]$ direction.
As such, the canting can be regarded as a consequence of the broken $C_3$ symmetry.

As discussed in the previous section, quantum order-by-disorder selects a discrete subset
of states out of the classical $U(1)$ degenerate manifold.
Our result strongly suggests that the NP phase originates from the FF phase
upon the incorporation of quantum fluctuations.
Furthermore, the NP phase is much enhanced in the quantum 
model as compared to the FF phase in the classical model, which we argue as follows:
(a) NP has an increased critical field of the transition into P relative to its classical counterpart.
(b) NP supersedes many of the classical long-range ordered states
\cite{chern_magnetic_2020,lee_magneticfield_2019} 
in a wide region of the phase diagram with ZZ being the only exception.

The transition from P into NP appears to be continuous with a gap 
closing and a consecutive range of fields with long correlation length. 
For Hamiltonians with only local interactions,
a long correlation length implies a small spectral gap \cite{hastings_locality_2004}.
Here we expect a small spectral gap from the quantum order-by-disorder scenario. 
In order to verify our hypothesis, we slightly tilt the magnetic field by 
$1^{\circ}$ from, and rotate it around, the $[111]$ axis. Within the range of fields
with a small gap, we find that the canting direction continuously follows the
rotated field with only small and smooth variation in the ground state energy (see Appendix \ref{app:u1signature}). The magnitude of the variation is of similar order as in LSWT. 

As the field is tuned, the fully quantum model exhibits transitions between different NP states at certain ratios $\Gamma/|K|$
similar to the transition between NP$_\gamma$ and NP$_{\alpha\beta}$ observed with LSWT in Sec.~\ref{scn:semiclassicalanalysis}.
Contrary to LSWT and an earlier tensor network calculation \cite{lee_magneticfield_2019}, however,
iDMRG reveal multiple transitions between NP$_\gamma$, NP$_{\alpha\beta}$, and other NP states.
The order in which the sets appear, as well as the critical fields separating them,
depend on the geometry used with iDMRG.
This is in stark contrast to the critical fields of transitions from P to NP and from NP to ZZ
which are relatively insensitive to the geometry.
As a consequence different NP states compete in a wide range in parameter space
and, therefore, we do not distinguish between them in the quantum phase diagram Fig.~\ref{fig:pd}.
In Sec.~\ref{subscn:nporderparameter} we introduce an order parameter for NP and discuss the different NP states in more detail.

A ZZ phase at intermediate fields exists embedded within NP for $\phi/\pi \gtrsim 0.25$, or 
$\Gamma/\lvert K \rvert \gtrsim 1$.
The upper transition between NP and ZZ is likely continuous.
Further lowering the field, the transition from ZZ back to NP appears to be first order 
and its critical field depends on the geometry used.
As a consequence, it is difficult to draw a firm conclusion on the extent of the ZZ phase.
While there is no evidence within iDMRG for a long-range magnetically ordered phase
at zero field for all the geometries (up to $L_\mathrm{circ}=12$ sites) explored here,
scenarios for the two-dimensional limit in which either the ZZ phase ranges down to zero field \cite{wang_one_2019a}
or other magnetic orderings \cite{lee_magneticfield_2019} are stabilized cannot be excluded.
In fact, various magnetic orders that appear in the classical model \eqref{eqn:Ham}
are obtained as meta-stable states at small $\chi$ and, thus,
these orderings are competing with NP in the quantum model.
We refer interested readers to Sec.~\ref{subscn:zerofield} for more details.

Upon the inclusion of a small $\Gamma'=-0.06$, the ZZ phase is 
stabilized at small fields.
The KSL shrinks and borders NP, ZZ and P.
The critical field of the transition from P into NP is slightly renormalized
to a smaller value due to the $\Gamma'$ interaction, as in the classical model.
Beyond that, the phenomenology of the phase transitions remains the same
as in the case of $\Gamma'=0$, namely the transition between P and NP
as well as that between NP and ZZ appear to be continuous.

\subsection{Order Parameter of the Nematic Paramagnet} \label{subscn:nporderparameter}

While the canting of the magnetic moments is an indicator of the broken $C_3$ symmetry, 
it is not a suitable order parameter in the zero field limit as the magnetic moments vanish.
A more adequate order parameter can be defined in terms of bond energies
$E_\gamma = \langle \mathcal H \rangle_{\langle i,j \rangle_\gamma}$, where $\gamma \in \{x,y,z\}$ is the label of the bonds.
The broken $C_3$ symmetry manifests in different $E_\gamma$ on each bond,
while if the $C_3$ symmetry is not broken, e.g. at high fields, the bond energies satisfy 
$E_x = E_y = E_z$. 
This motivates to define an order parameter $\mathcal O^\gamma_{C_3}$ on each bond $\gamma$
by the difference of $E_\gamma$ to the average bond energy as
\begin{eqnarray}
    \mathcal O_{C_3}^x &= E_x - \frac{1}{3} \left(E_x + E_y + E_z \right), \nonumber  \\ 
    \mathcal O_{C_3}^y &= E_y - \frac{1}{3} \left(E_x + E_y + E_z \right),  \\
    \mathcal O_{C_3}^z &= E_z - \frac{1}{3} \left(E_x + E_y + E_z \right). \nonumber
\end{eqnarray}

In terms of bond energies, NP$_\gamma$ is characterized by a single preferred bond,
$E_\gamma < E_\alpha \simeq E_\beta$ and hence
$\mathcal O^\gamma_{C_3} < 0 < \mathcal O^\alpha_{C_3} = \mathcal O^\beta_{C_3}$,
while for NP$_{\alpha\beta}$ two bonds are equally preferred,
$E_\alpha \simeq E_\beta < E_\gamma$ and hence
$\mathcal O^\alpha_{C_3} = \mathcal O^\beta_{C_3} < 0 < \mathcal O^\gamma_{C_3}$.
An arbitrary NP state has three different bond energies, $E_\alpha \neq E_\beta \neq E_\gamma$, and hence 
$\mathcal O^\alpha_{C_3} \neq \mathcal O^\beta_{C_3} \neq \mathcal O^\gamma_{C_3}$.

\begin{figure}
    \includegraphics[width=0.95\linewidth]{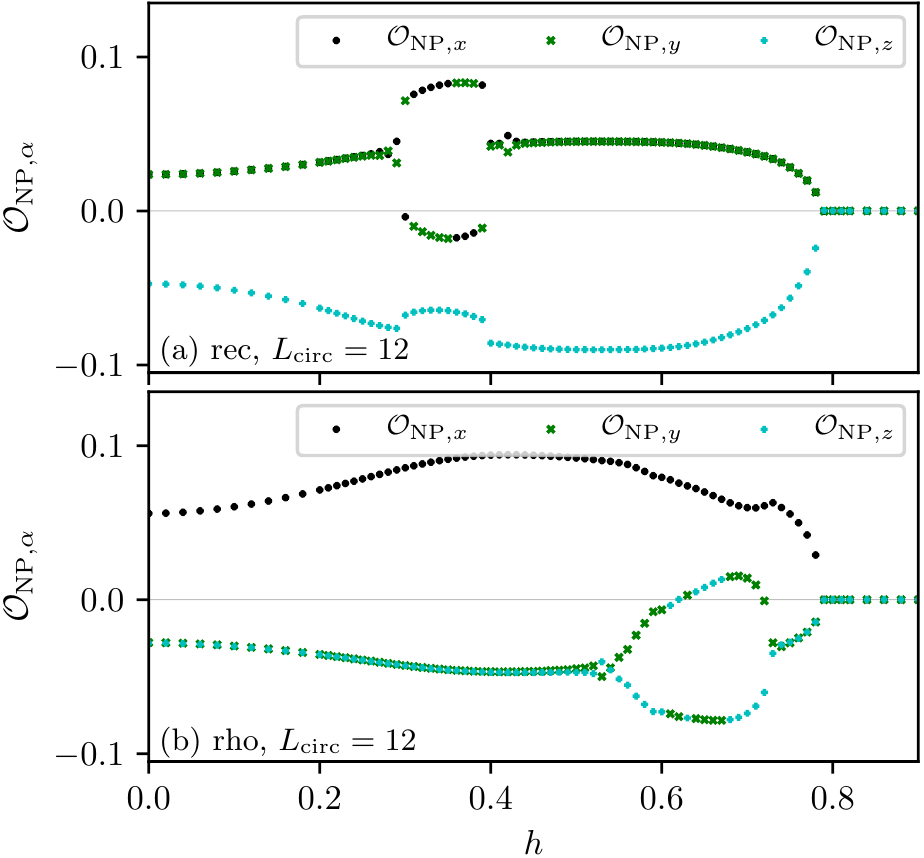}
    \caption{
        Order parameters $\mathcal O^\gamma_{C_3}$ for each bond $\gamma \in \{x,y,z\}$
        at $\phi/\pi=0.1$ on (a) the rectangular geometry with $L_\mathrm{circ}=12$
        and (b) the rhombic geometry with $L_\mathrm{circ}=12$.
        The non-zero $\mathcal O^\gamma_{C_3}$ indicates the broken $C_3$ symmetry of NP.
        Depending on the geometry both sets of nematic states occur.
        Namely, (a) the rectangular geometry selects NP$_\gamma$,
        while (b) the rhombic geometry selects NP$_{\alpha\beta}$ in a wide range of fields.
        Both geometries exhibit an intermediate NP phase with states  
        which neither correspond to NP$_\gamma$ nor NP$_{\alpha\beta}$.
    }
    \label{fig:O_NP}
\end{figure}

In any case, a finite $\mathcal O^\gamma_{C_3}$ indicates 
that the $C_3$ symmetry is broken, in particular as soon as the NP phase is stabilized (see Fig.~\ref{fig:O_NP}).
The $C_3$ symmetry remains broken down to the zero field limit.
Whether NP$_\gamma$, NP$_{\alpha\beta}$, or a different NP state gets selected is,
however, dependent on the geometry used in iDMRG.
While the rhombic geometry prefers NP$_{\alpha\beta}$ near the upper critical field
and in the zero field limit, the rectangular geometry prefers the NP$_\gamma$ phase.
Moreover, both geometries have in common an intermediate transition into a phase,
which exhibits the homogeneous canting of the magnetic moments characteristic for NP, 
but the azimuthal angle $\varphi$ neither belongs to NP$_\gamma$ nor NP$_{\alpha\beta}$.
Instead, degenerate states with $\varphi = n \pi/3 \pm \delta$ occur,
where $n$ is an odd (even) integer corresponding to NP$_\gamma$ (NP$_{\alpha\beta}$).
In the two-dimensional limit, this $\varphi$ implies a six-fold degeneracy.  

In conclusion, the different sets of NP states compete over a wide range of fields,
and their difference in energy appears to be of the order of,
or smaller than, finite-size corrections. 
Whether the intermediate NP state is an effect of quantum order-by-disorder
or of the finite circumference has to remain an open question.
In that context one should recall that the quantum order-by-disorder computation
which we present in Sec.~\ref{scn:semiclassicalanalysis} is semi-classical. 
Non-linear terms, which were not included, may be relevant to stabilize an intermediate NP phase
different from NP$_\gamma$ or NP$_{\alpha\beta}$ and with a six-fold degeneracy.

\subsection{Zero Field Limit of the $K \Gamma$ Model} \label{subscn:zerofield}
\begin{figure} 
    \includegraphics[width=0.95\linewidth]{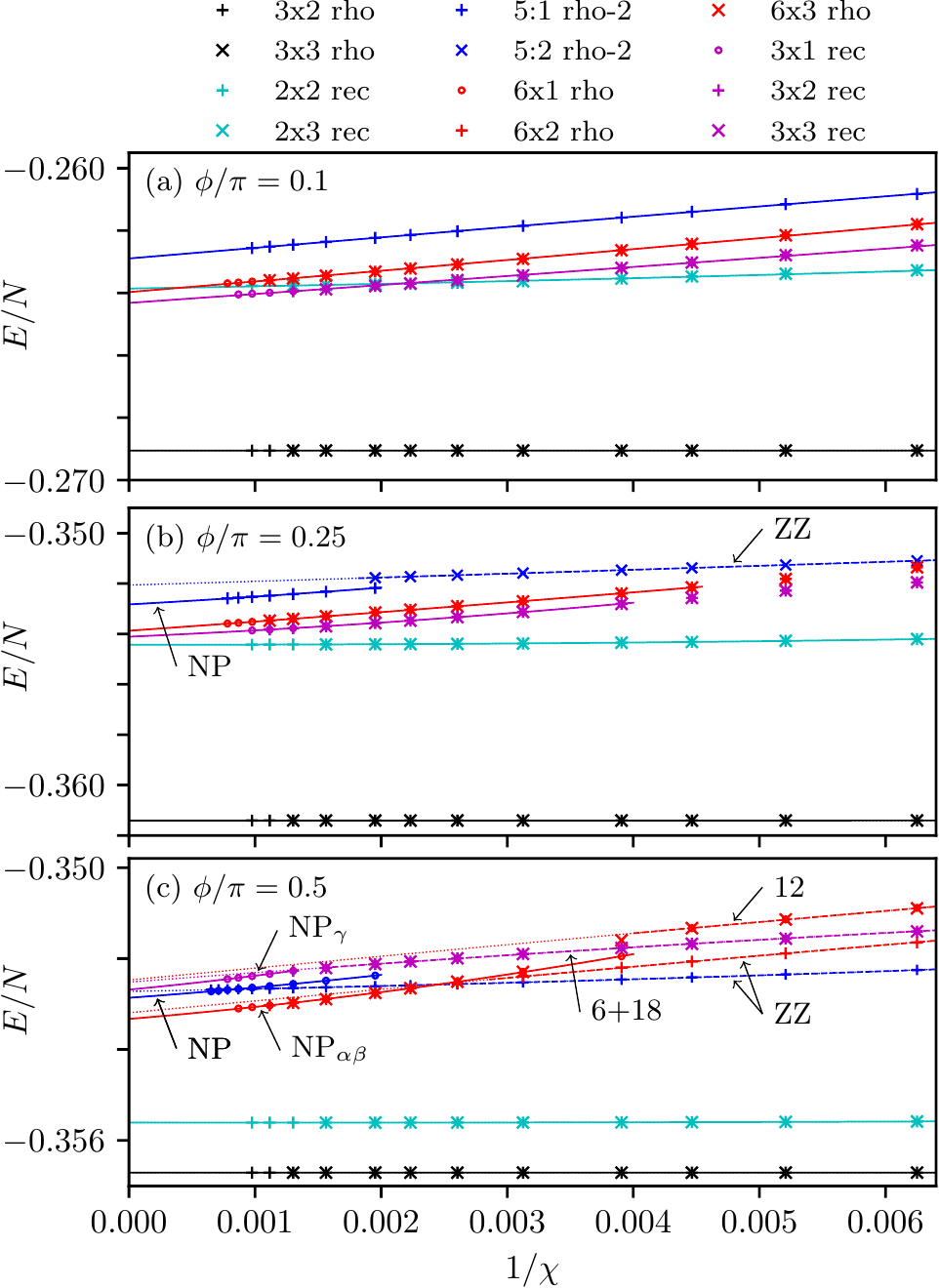}
    \caption{
            In the limit $h \longrightarrow 0$, the quantum ground state
            obtained by iDMRG depends on the cylinder geometry and the bond dimension $\chi$ of the MPS.
            The solid line represents a quadratic fit to the energies of MPS that belong to NP,
            while for dashed lines the MPS exhibits a long-ranged magnetically ordered state. 
            More specifically, we find the following states. 
            At (a) $\phi/\pi=0.1$ and (b) $\phi/\pi=0.25$ the ground state is either
            NP$_{\alpha\beta}$ for rhombic unit cells (rho) 
            or NP$_\gamma$ for rectangular unit cells (rec).
            The \emph{rhombic-2} (rho-2) geometry cannot be associated with either of them,
            as it does not exhibit the corresponding anisotropy in bond energies,
            namely $E_\alpha \simeq E_\beta < E_\gamma$ (NP$_{\alpha\beta}$)
            or $E_\gamma < E_\alpha \simeq E_\beta$ (NP$_\gamma$) with $a,b,c \in \{x,y,z\}$.
            Zigzag (ZZ) order is found close in energy for \emph{rhombic-2}.
            (c) In the pure $\Gamma > 0$ model, i.e. $\phi/\pi=0.5$, cylinders with 
            $L_\mathrm{circ} = 12$ (3x$L_y$ rec, and 6x$L_y$ rho) exhibit
            transitions from states with
            finite magnetic moment and long-range order at small $\chi$ 
            into either of the NP phase at large $\chi$.
            In particular, 
            \emph{5:2 rho-2} exhibits a ZZ order for $\chi \leq 1200$,
            as does \emph{6x2 rho} for $\chi \leq 450$,
            \emph{6x3 rho} exhibits a 12-site order for $\chi \leq 256$,
            \emph{3x1 rec} and \emph{3x2 rec} show a 6-site order for $\chi \leq 640$ 
            \emph{3x3 rec} exhibits an 18-site order for $\chi \leq 640$.
            This indicates that NP$_\gamma$, NP$_{\alpha\beta}$
            and several magnetically ordered states
            observed in the classical model \cite{chern_magnetic_2020} are close in energy.
        }
    \label{fig:EvsChi_h0}
\end{figure}

The ground state of the $K \Gamma$ model (i.e.~\eqref{eqn:Ham} with $\Gamma'=0$) at zero field
in the range $0 \le \phi/\pi \le 0.5$ has been a subject of debate recently.
Various numerical methods provide different answers
\cite{lee_magneticfield_2019,gohlke_quantum_2018,wang_one_2019a,luo_gapless_2019,wang_multinode_2020}.
Here, we use different cylinder geometries with circumferences as large as $L_\mathrm{circ}=12$
and vary the bond dimension $\chi$ of the matrix product state (MPS),
which encodes the quantum wave function,
to further support our interpretation that NP extends down to zero field.
However, as we discuss in the following, the finite circumference affects the ground state energy significantly
and a firm conclusion regarding the ground state in the two-dimensional limit cannot be drawn.

In Fig.~\ref{fig:EvsChi_h0}, we compare the ground state energies $E_{GS}$ of different geometries upon varying $\chi$ at $\phi/\pi = 0.1, 0.25$, and $0.5$.
Narrow cylinders with $L_\mathrm{circ} = 6$ (3x$L_y$ rho) or $L_\mathrm{circ}=8$ (2x$L_y$ rec)
show good convergence with respect to $\chi$ and feature an almost flat evolution of $E_{GS}$.
In contrast, cylinders with $L_\mathrm{circ}=10$ or $12$ require relatively large $\chi$ for convergence.
Nonetheless, we can read off the following tendencies.
Firstly, the rhombic geometry (rho) exhibits an NP$_{\alpha\beta}$ ground state,
while the rectangular geometry (rec) exhibits NP$_\gamma$ at sufficiently large $\chi$.
Secondly, $E_{GS}$ depends significantly on $L_\mathrm{circ}$
where the narrowest cylinder (3x$L_y$ rho) has the lowest energy,
which is about $2$ to $3\%$ below $E_{GS}$ of cylinders with $L_\mathrm{circ} = 12$.
This implies that finite-size effects are significant as will be discussed below.
Thirdly, some of the magnetically ordered states observed in the corresponding classical model \cite{chern_magnetic_2020} appear as meta-stable ground states
if an MPS ansatz with small $\chi$ is used,
in particular at $\phi/\pi = 0.5$, see Fig.~\ref{fig:EvsChi_h0} (c).
The difference in energy between the magnetically ordered states and the NP phase is much smaller
than that between different circumferences.
Consequently, finite-size effects make it difficult to conclude the precise nature of the ground state
near the $\Gamma$ limit as $h \longrightarrow 0$.

Remarkably, we find NP$_{\alpha\beta}$ to be adiabatically connected to uncoupled $K \Gamma$ chains \cite{yang_phase_2019}
along alternating $\alpha$ and $\beta$ bonds
\footnote{Likewise, we find NP$_\gamma$ to be adiabatically connected to uncoupled $K \Gamma$ dimers on bonds with label $\gamma$. This is consistent with a recent series expansion on a dimer ansatz \cite{yamada_ground_2020}, which finds the dimerized ansatz to be stable up to almost isotropic couplings.}.
That is to say, we do not observe a phase transition
--- other than a change to a different NP$_{\alpha\beta}$ orientation --- 
upon adiabatically turning off the couplings on all bonds $\langle i,j \rangle_\gamma$ with a label $\gamma \in \{x,y,z\}$.
In particular, consider the rhombic geometry with bond labels as in Fig.~\ref{fig:geom}(a).
The ground state obtained by iDMRG is NP$_{yz}$ with $E_y = E_z < E_x$.
We denote the interactions on the $x$ bond collectively as $J_x$,
and multiply it by the factor $1-\eta$ with $\eta \in [0,1]$.
In doing so, we do not find a phase transition upon tuning $\eta$ from $0$ to $1$.
On the other hand, if the interactions on the $y$ or $z$ bonds are switched off
using a similar parametrization, a different orientation, NP$_{xz}$ or NP$_{xy}$ respectively,
is selected at a small but finite $\eta$.

The three possible orientations of the NP$_{\alpha\beta}$ are exactly degenerate
in the two-dimensional limit.
This degeneracy, however, is lifted on the cylinder due to its finite circumference,
which is best understood in the limit of the uncoupled $K \Gamma$ chains.
On the rhombic geometry, see Fig.~\ref{fig:geom}(a), NP$_{yz}$ corresponds to a $K \Gamma$ chain
composed of alternating $y$ and $z$ bonds along the circumference.
Such a chain is finite with $L_\mathrm{circ}$ sites and subjected to periodic boundary conditions.
The remaining two orientations instead correspond to infinite chains.
The ground state energies obey $E|_{L=6} < E|_{L=12} < ... < E|_{L=\infty}$.
Here $E_{L=6}$ is about $1$ to $2\%$ smaller than that of $L=\infty$, which is of the same energy scale as over which $E_{GS}$ spreads for the different geometries, see Fig.~\ref{fig:EvsChi_h0}.
Thus, the significant finite-circumference --- or finite-size --- effects of the two-dimensional $K \Gamma$ model can be explained,
at least within the NP$_{\alpha\beta}$ set of the NP phase,
by its relation to the $K \Gamma$ chains. 
We expect this type of finite-size effect to be quite ubiquitous in any numerical study
of an extended spin-1/2 Kitaev model with the Kitaev and other anisotropic exchange interactions 
using a finite geometry
\cite{gohlke_quantum_2018,gordon_theory_2019,luo_gapless_2019,wang_one_2019a,wang_pressure_2019,wang_multinode_2020}.

%%%%%%%%%%%%%%%%%%%%%%%%%%%%%%%%%%%%%%%%%%%%%%%%%%%%%%%%%%%%%%%%%%%%%%%%%%%%%%%%%%%%%%%%%%
%%% Dynamics
%%%%%%%%%%%%%%%%%%%%%%%%%%%%%%%%%%%%%%%%%%%%%%%%%%%%%%%%%%%%%%%%%%%%%%%%%%%%%%%%%%%%%%%%%%
\section{Dynamics} \label{scn:dyn}
\begin{figure*}[tb]
    \includegraphics[width=\linewidth]{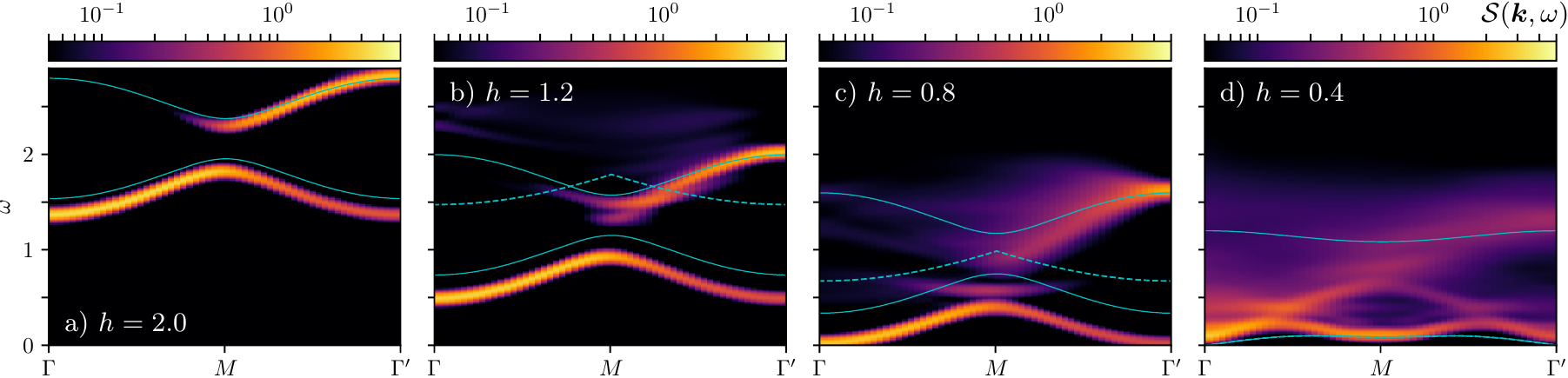}
    \caption{
        Dynamical spin structure factor $\mathcal S(\mathbf{k},\omega)$ at $\phi/\pi=0.1$.
        (a) $h=2.0$ and (b) $h=1.2$ in the high-field paramagnet (P) phase,
        (c) $h=0.8$ in P right before the transition into the nematic paramagnet (NP),
        and (d) $h=0.4$ within NP.
        In (a)-(c) P exhibits two magnon bands, 
        with the lower one remaining sharp but the upper one being obscured by the multi-magnon continuum upon decreasing the field.
        Comparing to the dispersion obtained by LSWT,
        the lower magnon band (solid line) is shifted downwards in energy and
        hits zero at the $\Gamma$ and $\Gamma'$ point
        at a higher critical field than that predicted classically.
        Within NP, as in (d), the lower magnon band becomes slightly diffusive.
        A broad continuum appears, ranging from (almost) zero frequency to about $\omega=1.6$.
        The spectral weight is mostly concentrated at low frequencies near the $\Gamma$-point,
        indicating long wavelength excitations corresponding to a pseudo-Goldstone mode.
        LSWT spectrum in the FF phase, or NP in the quantum limit, is gapless by the Nambu-Goldstone theorem.
        Some of the spectral weight is redistributed to the $M$-point,
        signaling that a small perturbation, e.g. $\Gamma'$, may be sufficient
        to drive the system into the zigzag (ZZ) order.
        }
    \label{fig:dsf}
\end{figure*}
The \emph{dynamical spin structure factor} $\mathcal S(\mathbf{k},\omega)$
contains information about the excitation spectrum and can be probed by inelastic neutron scattering experiments.
We consider 
$\mathcal S(\mathbf{k},\omega) = \sum_{\gamma \in \{x,y,z\}} \mathcal S^{\gamma\gamma} (\mathbf{k},\omega)$
with
$\mathcal S^{\gamma\gamma} (\mathbf{k},\omega)$ being the spatio-temporal Fourier transform
of the dynamical correlations 
\begin{equation}
    \mathcal S^{\gamma\gamma} (\mathbf{k},\omega) = N \int \mathrm{d}t ~ e^{i \omega t} \sum_{a,b} e^{i \mathbf{k} \cdot (\mathbf{r}_b - \mathbf{r}_a)} ~ C^{\gamma\gamma}_{ab}(t) ~,
\end{equation}
where $\gamma \in {x,y,z}$ is the spin component,
$\mathbf{r}_a$ and $\mathbf{r}_b$ are the spatial positions of the spins,
$N$ is a normalization factor defined via 
$\int d\omega \int \mathrm{d} \mathbf{k} ~ \mathcal S^{\gamma\gamma} ( \mathbf{k},\omega) = \int \mathrm{d} \mathbf{k}$,
and $C^{\gamma \gamma}_{ab} (t)$ denotes the dynamical spin-spin correlation
$C_{ab}^{\gamma \gamma} (t) = \langle \psi_0 | S_a^\gamma (t) S_b^\gamma (0) | \psi_0 \rangle$.
Given the ground state wave function in MPS form,
the time evolution is carried out using the tMPO method \cite{zaletel_timeevolving_2015}
until $t = 60$.
The time series is then extended by a linear prediction \cite{yule_vii_1927,makhoul_linear_1975,barthel_spectral_2009} and multiplied with a Gaussian distribution.
The resulting line broadening amounts to $\sigma_\omega = 0.036$.
We restrict ourselves to magnetic fields near the transition between P and NP,
where the numerical errors and finite size effects are found to be small.
A rhombic geometry with $L_\mathrm{circ} = 6$ sites is used such that
three separated lines of accessible momenta exist in the first Brillouin zone,
as illustrated in Fig.~\ref{fig:geom}.

We present two sets of $\mathcal S(\mathbf{k},\omega)$,
Fig.~\ref{fig:pd}(c-f) and Fig.~\ref{fig:dsf},
along the high-symmetry direction $\Gamma-M-\Gamma'$ in the Brillouin zone.
In Fig.~\ref{fig:pd}(c-f) we want to emphasize the similarities and differences of
$\mathcal S(\mathbf{k},\omega)$ in the two limits (c,d) $\Gamma \longrightarrow 0$
and (e,f) $K \longrightarrow 0$.
Whereas in Fig.~\ref{fig:dsf} we focus on the evolution of the magnon bands
upon reducing the field across the critical field while keeping $\phi/\pi=0.1$ fixed.
$\mathcal S(\mathbf{k},\omega)$ for additional parameters are presented in Appendix \ref{app:additionaldsf}.

As highlighted before, the NP phase remains stable in a wide range in parameter space.
NP extends from the almost (FM) Kitaev limit, $\phi \longrightarrow 0$,
to the (AF) $\Gamma$ limit, $\phi/\pi \longrightarrow 0.5$ and beyond when $\Gamma'=0$.
In the entire range, the lower magnon band which is apparent in P reduces in energy
upon approaching the critical field of the transition P to NP,
where eventually it condenses at $\Gamma$ and $\Gamma'$.
Within NP, the main spectral weight is observed at long wavelengths and low frequencies
above a small spectral gap.
Such excitations are consistent with the slightly lifted accidental $U(1)$ degeneracy
and correspond to the slow pseudo-Goldstone mode and is the characteristic feature of NP.

On the contrary, the continuum exhibits a very distinct structure in both limits of $\phi$.
Near $\phi \longrightarrow 0$, that is near the pure Kitaev model with $K<0$,
$\mathcal S(\mathbf{k},\omega)$ bears similarities to that of the nearby ferromagnetic KSL.
The broad continuum of the KSL is formed by itinerant Majorana fermions
subject to a quantum quench caused by the excitation of a flux pair
when acting with a spin operator on the ground state
\cite{knolle_dynamics_2014,knolle_dynamics_2015}.
Since the flux-pair excitation is gapped, $\mathcal S(\mathbf{k}, \omega)$ has a finite gap
even though the Majorana fermion spectrum is gapless.
The continuum above the spectral gap has an almost $\mathbf k$-independent width equal to that
of the single Majorana fermion dispersion.
$\mathcal S(\mathbf{k},\omega)$ of NP and P proximate to the KSL exhibit a similar continuum 
with the same upper cutoff frequency $\omega \approx 1.6$.
Thus, remnants of the Majorana fermion excitations of the KSL appear to persist at higher energies.
Differences between NP and KSL are most obvious at lower energies.
While the KSL has a relatively low-lying dispersion and the spectral gap,
NP exhibits a somewhat broadened dispersing low-energy mode
associated with the pseudo-Goldstone mode.
As a consequence, it may be challenging to discern NP and the nearby KSL
using inelastic neutron scattering experiments.
Such a phenomenology, which is known as \emph{proximate spin liquid} \cite{banerjee_proximate_2016},
was previously observed in the Kitaev-Heisenberg model \cite{yamaji_clues_2016,gohlke_dynamics_2017}.

In the other limit $\phi/\pi \longrightarrow 0.5$, i.e. the pure $\Gamma$ model with $\Gamma>0$,
$\mathcal S(\mathbf k, \omega)$ of NP at $h=1.4$ is dominated by diffuse remnants of the magnon modes that exist in P at $h=1.6$, compare Fig.~\ref{fig:pd}(e) and (f). 
The magnon mode has an enhanced bandwidth and dispersion which remains of equal magnitude upon approaching the transition.
This is in contrast to the pure Kitaev model in a $[111]$ field, which exhibits a band that simultaneously flattens and approaches zero in the entire reciprocal space \cite{gohlke_dynamical_2018}.
Moreover, the continuum within NP has more structure and an increased energy scale, $\Delta\omega/\Gamma \approx 2.5$, compared to that of the KSL, $\Delta\omega/K \approx 1.6$.
In comparison to the $\Gamma$ model (with $\Gamma>0$) at zero field \cite{samarakoon_classical_2018}, 
we observe a similar energy scale of the continuum.
At low energies, however, the main spectral weight shifts from $M$ at zero field,
to $\Gamma$ at high field, which corresponds to the pseudo-Goldstone mode.
Above the pseudo-Goldstone mode, in particular at $\Gamma$ and $\Gamma'$, the continuum exhibits a gap similar to the gap observed at zero field.

Figure~\ref{fig:dsf} demonstrates the evolution of the magnon modes
upon approaching the critical field.
Two magnon bands exists due to the unit cell having two sites.
The two magnon modes are most pronounced at high fields, e.g.~Fig.~\ref{fig:dsf}(a) at $h=2.0$,
where the magnetic moment is close to saturation.
Agreement with LSWT is good except from an overall shift
of the lower magnon band towards smaller frequencies.
Upon reducing the field, e.g.~Fig.~\ref{fig:dsf}(b) at $h=1.2$,
both magnon bands move downwards in energy,
while their shapes are essentially the same as those at $h=2.0$.
However, the upper band starts to overlap with the two-magnon continuum resulting
in a broader linewidth due to additional decay channels.
The bottom of the two-magnon continuum as obtained from LSWT is marked by the dashed line.
In the fully quantum computation (tMPO), however, the continuum starts at lower frequencies
due to the overall reduction of the lower magnon band as compared to LSWT.

Right before entering NP, e.g.~Fig.~\ref{fig:dsf}(c) at $h=0.8$,
the lower magnon band approaches zero frequency at the $\Gamma$ and $\Gamma'$ points,
illustrating a condensation of magnons and the imminent uniform canting 
of the magnetic moments.
LSWT predicts a gap, as the classical model is still in the fully-polarized phase due to a lower $h_\mathrm{crit}$.
The continuum overlaps with the upper magnon band and obscures it except near the $\Gamma'$ point. 
Upon the transition from P to NP, the continuum does not change significantly
and, thus, confirms that the transition is continuous. 

Within NP, e.g.~Fig.~\ref{fig:dsf}(d) at $h=0.4$,
a broad continuum appears and extends up to $\omega \approx 1.6$.
Furthermore, a softening of the DSF at the M point signals the development
of a significant ZZ correlation when the magnetic field is lowered.
Adding a small $\Gamma'<0$ indeed leads to the ZZ order as in Fig.~\ref{fig:pd}(b).
The ZZ correlation may be related to the presence of an intermediate ZZ order
that takes place at larger $\Gamma/\lvert K \rvert$ ($\phi/\pi \gtrsim 0.25$)
in the quantum phase diagram Fig.~\ref{fig:pd}(a).
In the corresponding classical model, the intermediate ZZ order extends down to
$\phi/\pi = 0.1$ \cite{chern_magnetic_2020}.
Thus, the ZZ correlation may also be a remnant of the classical ZZ order.
The LSWT spectrum is gapless at the $\Gamma$ point as the classical model is now in the FF phase.
Analogous to tMPO, LSWT predicts a condensation of magnons at the $M$ point,
which becomes more pronounced upon further lowering the field
and approaching the transition into ZZ in the classical model
(see Appendix \ref{app:additionallswt}).

%%%%%%%%%%%%%%%%%%%%%%%%%%%%%%%%%%%%%%%%%%%%%%%%%%%%%%%%%%%%%%%%%%%%%%%%%%%%%%%%%%%%%%%%%%
%%% Discussion/Conclusion
%%%%%%%%%%%%%%%%%%%%%%%%%%%%%%%%%%%%%%%%%%%%%%%%%%%%%%%%%%%%%%%%%%%%%%%%%%%%%%%%%%%%%%%%%%
\section{Discussion} \label{scn:con}

We demonstrate that the $K \Gamma$ and $K \Gamma \Gamma'$ models in a magnetic field
along the $[111]$ direction support an NP phase,
which is not magnetically ordered, but breaks the lattice-rotation symmetry
while preserving translational symmetry.
We trace the origin of NP to the FF phase
in the corresponding classical model, which has a $U(1)$ degenerate manifold of states. 
When zero-point quantum fluctuations are incorporated,
the $U(1)$ degeneracy is broken down to a discrete subset of ground states 
related by the $C_3$ symmetry.
This quantum order-by-disorder effect leads to a pseudo-Goldstone mode \cite{rau_pseudogoldstone_2018}
in the excitation spectrum, as shown in the dynamical spin structure factors of the quantum model.

In realistic Kitaev materials like \arucl, the lattice-rotational symmetry
has already been broken by a monoclinic distortion \cite{johnson_monoclinic_2015,kim_crystal_2016}.
Therefore, it is reasonable to expect that the degeneracy of NP is lifted,
i.e.~one or two out of the six canting directions is favored.
In fact, a substantial magnetic torque is measured for \arucl~at high fields \cite{leahy_anomalous_2017},
indicating that NP may be stabilized in \arucl.
While monoclinic distortion may have already caused a small canting of the magnetic moment,
we argue that the canting of magnetic moments in NP is much more pronounced.
It would be necessary to look for a second transition at high fields
to verify the existence of the field-induced NP,
which is sandwiched between the long-range zigzag order and the longitudinally-polarized paramagnet.
The latter only exhibits a negligible torque,
while NP has a much larger torque due to the induced canting.
The specific nature of the distortion may either enhance or mitigate the extent of NP.

Considering the wide range in parameter space over which NP occurs,
this phase may as well be relevant to other Kitaev materials.
In the context of the NP phase, with its spontaenously broken C$_3$ symmetry,
a particularly interesting example is the iridate K$_2$IrO$_3$.
Recently, K$_2$IrO$_3$ was proposed to exhibit ferromagnetic Kitaev exchange and large off-diagonal exchange while maintaining a C$_3$ point group symmetry \cite{yadav_large_2019}.

An important implication for experiments is drawn
from the dynamical spin structure factor of the NP phase,
which exhibits diffusive scattering features masking sharp magnonic excitations.
Such a scattering continuum bears some similarities to that of the nearby KSL.
With a greater ratio of $\Gamma/|K|$, the continuum appears in a wider range of energies.
Therefore, the excitation continuum observed in inelastic neutron scattering experiments
on \arucl \cite{banerjee_proximate_2016,banerjee_neutron_2017,winter_probing_2018,banerjee_excitations_2018,balz_finite_2019}
could originate from the NP phase.
The intriguing question of whether NP is a topologically non-trivial phase
is an excellent subject of future study.

%%%%%%%%%%%%%%%%%%%%%%%%%%%%%%%%%%%%%%%%%%%%%%%%%%%%%%%%%%%%%%%%%%%%%%%%%%%%%%%%%%%%%%%%%%
%%% Acknowledgements
%%%%%%%%%%%%%%%%%%%%%%%%%%%%%%%%%%%%%%%%%%%%%%%%%%%%%%%%%%%%%%%%%%%%%%%%%%%%%%%%%%%%%%%%%%
    \section{Acknowledgements} 
    
    We are grateful to R. Kaneko, K. Penc, J. G. Rau, J. Romhanyi, and N. Shannon for insightful discussions.
    M.G. acknowledges support by the Theory of Quantum Matter Unit of the Okinawa Institute of Science and Technology Graduate University (OIST)
    and by the Scientific Computing section of the Research Support Division at OIST for providing the HPC ressources.
    L.E.C. and Y.B.K. were supported by the Killam Research Fellowship from the Canada Council for the Arts, the NSERC of Canada, and the Center for Quantum Materials at the University of Toronto.
    HYK is supported by NSERC of Canada and the Center for Quantum Materials at the University of Toronto.
    This research was supported in part by the National Science Foundation under Grant No. NSF PHY-1748958.

%%%%%%%%%%%%%%%%%%%%%%%%%%%%%%%%%%%%%%%%%%%%%%%%%%%%%%%%%%%%%%%%%%%%%%%%%%%%%%%%%%%%%%%%%%
%%% Appendix
%%%%%%%%%%%%%%%%%%%%%%%%%%%%%%%%%%%%%%%%%%%%%%%%%%%%%%%%%%%%%%%%%%%%%%%%%%%%%%%%%%%%%%%%%%
\appendix
\section{Minimization of the Classical Energy in the Ferromagnetic Phase} \label{app:classicalminimization}
In this section, we work in the crystallographic ($abc$) coordinates, and write the spin 
components in \eqref{eqn:O3spin} as $\mathbf{S}=\left(S^a,S^b,S^c\right)$. 
For a ferromagnetic order, the Hamiltonian \eqref{eqn:Ham} reduces to 
\cite{chern_magnetic_2020}
\begin{equation} \label{FMhamiltonian}
\begin{aligned}[b]
H = & N \lbrace K \lvert \mathbf{S} \rvert^2 + \gamma \left[ - (S^a)^2 - (S^b)^2 + 2 
(S^c)^2 \right] \\
&- 2 (h^a S^a + h^b S^b + h^c S^c) \rbrace + \lambda \left( \lvert \mathbf{S} \rvert^2 - 
S^2 \right) ,
\end{aligned}
\end{equation}
where $N$ is the total number of unit cells, $\gamma \equiv \Gamma+2\Gamma'$ (assuming 
nonzero), and the Lagrange multiplier $\lambda$ has been introduced to constrain the spin 
magnitude. As any ferromagnetic order saturates the lower bound of the classical energy 
of the Kitaev interaction with $K<0$ \cite{baskaran_spins_2008}, we can simply drop it 
from \eqref{FMhamiltonian} in the subsequent analysis. Extremizing $H$ with respect to 
the variables $S^a$, $S^b$, $S^c$ and $\lambda$ leads to the following equations
\begin{subequations}
\begin{align}
& \frac{\partial H}{\partial S^a} = -2 \gamma S^a - 2 h^a + 2 \lambda S^a=0, 
\label{Saminimize} \\
& \frac{\partial H}{\partial S^b} = -2 \gamma S^b - 2 h^b + 2 \lambda S^b=0, 
\label{Sbminimize} \\
& \frac{\partial H}{\partial S^c} = 4 \gamma S^c - 2 h^c + 2 \lambda S^c=0, 
\label{Scminimize} \\
& \frac{\partial H}{\partial \lambda} = (S^a)^2+ (S^b)^2 + (S^c)^2 - S^2=0. 
\label{lambdaminimize}
\end{align}
\end{subequations}
We study the case where the field is completely aligned in the $[111]$ direction, i.e. 
$h^a=0$ and $h^b=0$, but $h^c \neq 0$. \eqref{Saminimize} becomes $-2 \gamma S^a + 2 
\lambda S^a=0$. \eqref{Sbminimize} yields a similar condition. \\ \\
\textbf{i.} Suppose that $S^a \neq 0$ \textit{or} $S^b \neq 0$. Then $\lambda=\gamma$. 
\eqref{Scminimize} then implies $S^c=h^c/3\gamma$, which is a physical solution only when 
$h^c/3\gamma \leq S$. \\ \\
\textbf{ii.} Suppose that $S^a=0$ \textit{and} $S^b=0$. Then, \eqref{Saminimize} and 
\eqref{Sbminimize} are satisfied. Futhermore, $\lambda \neq -2 \gamma$, otherwise 
\eqref{Scminimize} would imply $h^c=0$. Therefore, $S^c=h^c/(\lambda+2 \gamma)$, with 
$\lambda$ chosen to satisfy the normalization \eqref{lambdaminimize}. \\

In conclusion, the frustrated ferromagnet (FF) can be realized only when $h^c < 3 \gamma S$. As $h^c \geq 3 \gamma S$, the system becomes fully polarized. We thus identify $3 \gamma S$ as the critical field $h_\mathrm{crit}$.

\section{Signatures of the Accidental $U(1)$ Degeneracy} \label{app:u1signature}
\begin{figure}
    \includegraphics[width=\linewidth]{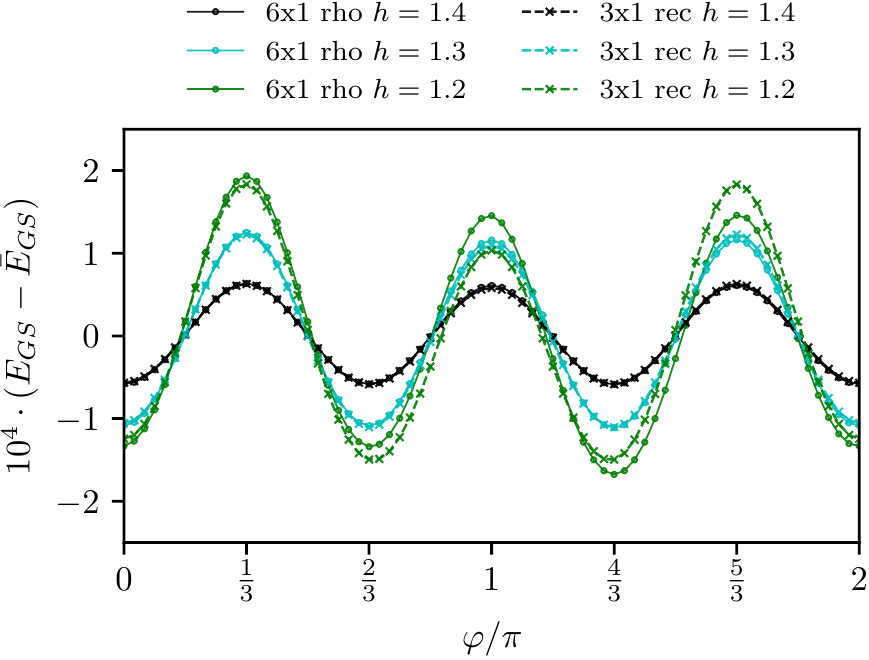}
    \caption{
        In order to verify the nearly degenerate $U(1)$ manifold of ground states in the quantum model \eqref{eqn:Ham},
        we tilt the magnetic field slightly from the $[111]$ axis by $\vartheta=1^\circ$ and then rotate it by varying $\varphi$.
        Plotted are the ground state energy as obtained with iDMRG for two different geometries
        with 12 sites circumference each, 
        \emph{6x1 rho} [see Fig.~\ref{fig:geom}(a)] and \emph{3x1 rec} [see Fig.~\ref{fig:geom}(b)].
        The average energy $\bar E_{GS} = 1/{2\pi} \int_{0}^{2\pi} d\varphi E_{GS}(\varphi)$
        has been subtracted.
        The magnetic moment follows the rotating field
        smoothly with a slight deviation up to $7^\circ$.
        As discussed in Sec.~\ref{scn:semiclassicalanalysis},
        the classical $U(1)$ degeneracy is lifted by quantum fluctuations. 
        The induced gap is of the order of $\varepsilon \approx 10^{-4}$
        and increases upon lowering the field.
        The minima are located at $\varphi = 0$, $\frac{2}{3}\pi$, and $\frac{4}{3}\pi$
        corresponding to NP$_{\alpha\beta}$.
    }
    \label{fig:E-Em_tilt}
\end{figure}
As shown in Sec.~\ref{scn:semiclassicalanalysis},
quantum order-by-disorder lifts the accidental $U(1)$ ground state degeneracy
in the classical model \eqref{eqn:Ham}.
In order to see whether this happens in the quantum model using iDMRG, we apply a magnetic field
\begin{equation}
\mathbf{h} = h \left( \sin \vartheta \cos \varphi \, \hat{\mathbf{a}} + \sin \vartheta \sin \varphi \, \hat{\mathbf{b}} + \cos \vartheta \, \hat{\mathbf{c}} \right)
\end{equation}
that is slightly tilted away from and rotated around the $[111]$ axis.
Here, we choose $\vartheta = 1^\circ$.
Upon rotation of the tilted field around the $[111]$ axis by varying $\varphi \in [0,2\pi)$,
we observe the following.
Firstly, the variation in $E(\vartheta=1^\circ, \varphi)$ is generally small, of the order $10^{-4}$.
The variation is the smallest near the P to NP transition and increases upon lowering the field.
Secondly, the tilting angle $\varphi_m$ of the magnetic moments continuously follows 
that of the field $\varphi_h$,
though there is a small deviation $\lvert \varphi_m - \varphi_h  \rvert \lesssim 7^\circ$.
Thirdly, $E(\vartheta=1^\circ,\varphi)$ has a period of $2/3\pi$.

The first observation is consistent with the quantum order-by-disorder scenario since quantum fluctuations only result in small corrections to the energy.
As a result, the gap induced by quantum order-by-disorder is small and can be easily overcome by small perturbations, e.g.~a slight tilting of the field.
While the second observation is a consequence of the first one,
it also implies that one can continuously tune from NP$_{\alpha\beta}$ to NP$_\gamma$
as well as between different orientations thereof.
Moreover, NP$_\gamma$ and NP$_{\alpha\beta}$ are related by the same spontaneous symmetry breaking.

The third observation is linked to the ground state degeneracy.
In NP the $C_3$ symmetry is broken spontaneously,
which is apparent from the anisotropic bond energies and the canting of the induced magnetic moment.
As a consequence, acting with a $C_3$ rotation
on the ground state transforms the state to a different state with the same energy.
This implies at least a threefold degeneracy which is manifest in the $2\pi/3$ periodicity in Fig.~\ref{fig:E-Em_tilt}.
Deviations from an exact threefold degeneracy are caused by the cylindrical geometry.
They are most pronounced for small $L_\mathrm{circ}$ and upon reducing the field.

In summary, our observations from tilting and rotating the magnetic field away from and around the $[111]$ axis are consistent with the scenario of NP$_\gamma$ and NP$_{\alpha\beta}$
being related to the classical FF phase.
Quantum fluctuations induce a small gap and only a discrete subset of states is selected out of the $U(1)$ manifold in the classical model.

\section{Additional Plots of Dynamical Spin Structure Factor} \label{app:additionaldsf}
\begin{figure*}[tb]
    \includegraphics[width=\linewidth]{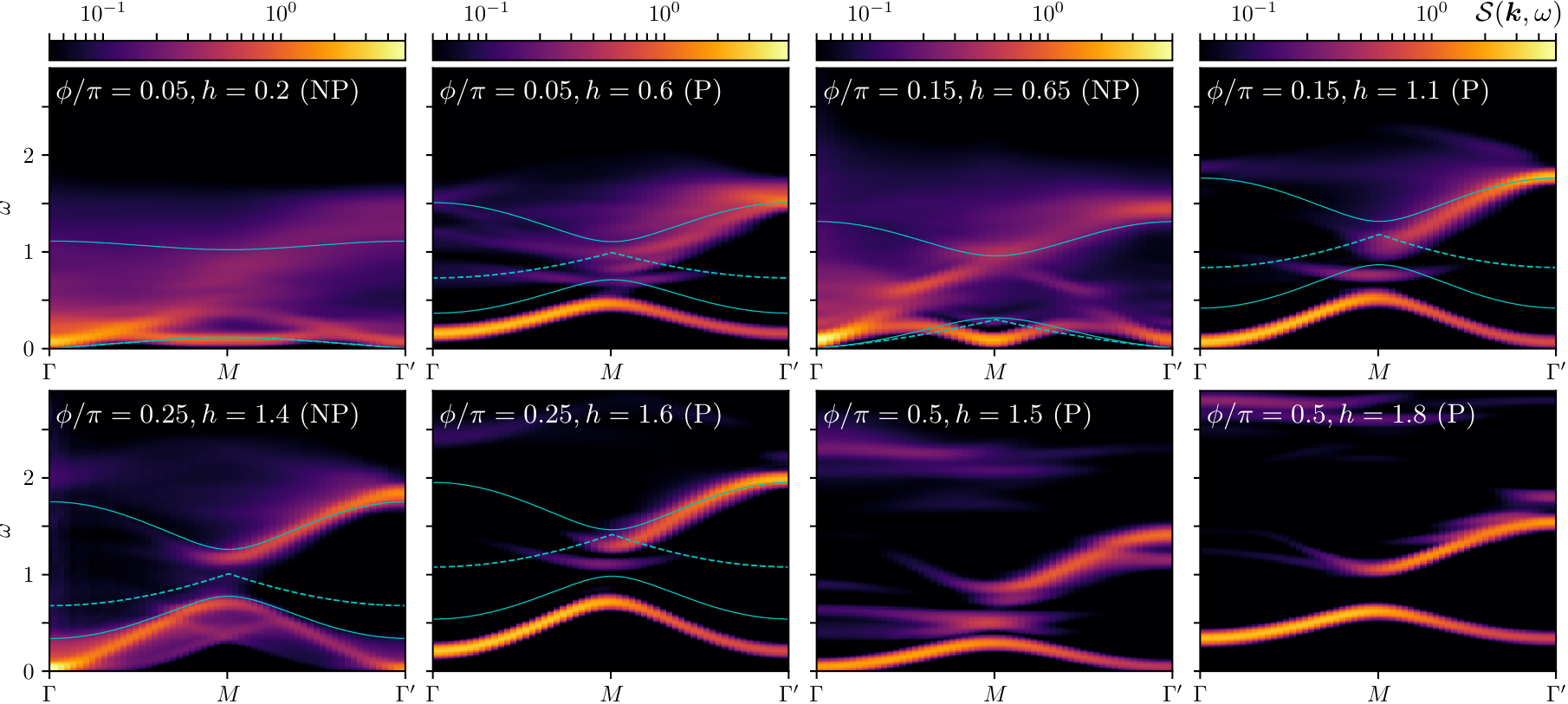}
    \caption{
        Dynamical spin structure factors $\mathcal S(\mathbf{k},\omega)$
        complementing Fig.~\ref{fig:pd}(c-f) and Fig.~\ref{fig:dsf} in the main text.
        Superimposed are the single-magnon bands (solid line)
        and the onset of the two-magnon continuum (dashed line) as obtained from LSWT.
        Near the Kitaev limit, $\phi/\pi=0.05$, there exists a broad continuum akin to the
        KSL and an almost flat band at low energies within P.
        Its dispersion increases upon tuning up $\phi$
        and reaches a maximum around $\phi/\pi = 0.25$.
        In the pure $\Gamma$ model, $\phi/\pi = 0.5$, the dispersion is still present although reduced.
        In the entire region $\phi/\pi = \left(0,0.5\right]$ the lower band attains its minimum
        at the Brillouin zone center, i.e.~the $\Gamma$ and $\Gamma'$ points,
        and reaches zero frequency at the transition between P and NP.
        Within NP the main spectral weight is located at low frequencies and long wavelengths,
        which is consistent with the existence of a pseudo-Goldstone mode.
        Some spectral weight is distributed around the M point,
        where the band bends towards lower energies. 
        Thus, a small perturbation, e.g.~$\Gamma'<0$,
        may trigger a transition into the ZZ order.
    }
    \label{fig:dsf_app}
\end{figure*}

We complement the discussion on dynamical spin structure factors $\mathcal S(\mathbf{k},\omega)$ by presenting the data for additional sets of parameters, as shown in Fig.~\ref{fig:dsf_app}.
We restrict ourselves to the phases close to the transition between P and NP,
where the numerical errors and finite size effects are found to be small.
Here we employ the rhombic geometry with $L_\mathrm{circ} = 6$ sites
that has three separated lines of accessible momenta in the first Brillouin zone,
as shown in Fig.~\ref{fig:geom} (a,b).

The P phase clearly exhibits two magnon bands, where the lower band attains its minimum at the Brillouin zone center, i.e. the $\Gamma$ and $\Gamma'$ points, for any $\phi/\pi \in \left(0,0.5\right]$. Upon approaching the transition, the lower band touches zero frequency at $\Gamma$ and $\Gamma'$, which is consistent with the uniform canting of spins in the NP phase.

Regardless of the specific value of $\phi$, $\mathcal S(\mathbf{k},\omega)$ of the NP phase features a nearly vanishing gap at $\Gamma$ that opens slightly upon further lowering the field.
The small gap gives rise to a multi-particle continuum near the zero energy,
partially obscuring the single-magnon mode by enabling decay channels.
The spectral weight is mostly concentrated just above the gap at the $\Gamma$,
signifying low-energy excitations with long wavelengths.
Such excitations correspond to the pseudo-Goldstone mode of the slightly lifted $U(1)$ degenerate manifold of states.
Moreover, $\mathcal S(\mathbf{k},\omega)$ exhibits a distinctive feature at high energies above $\Gamma'$,
which, as suggested by LSWT, appears to be a remnant of the upper branch of the one-magnon excitation.

Superimposed in Fig.~\ref{fig:dsf_app} are the two single-magnon bands and the resulting lower edge of the two-magnon continuum calculated by LSWT.
While in the P phase the bandwidth and the dispersion generally agree with the DSF computed by tMPO, 
the lower band of the latter is shifted towards lower energies.
Within NP (or FF in the classical model), LSWT exhibits a gapless excitation at $\Gamma$ and $\Gamma'$ that is the Nambu-Goldstone mode arising from spontaneously breaking a continuous symmetry.

In the limit $\phi \longrightarrow 0$ the system enters the KSL phase,
which is characterized by the fractionalization of spins into itinerant Majorana fermions with a background of static $\mathbb Z_2$ fluxes \cite{kitaev_anyons_2006}.
The fluxes are gapped excitations of an emergent $\mathbb Z_2$ gauge field.
$\mathcal S(\mathbf{k},\omega)$ of the KSL features a spectral gap $\Delta$ equal to the energy of a flux pair.
A continuum starts at $\Delta$ and has a width equal to that of the single Majorana fermion dispersion \cite{knolle_dynamics_2014,knolle_dynamics_2015}.
The continuum is almost entirely formed by Majorana fermions subject to a quantum quench 
that caused by exciting a flux pair \cite{knolle_dynamics_2014}.

In the NP or P phase proximate to the KSL, see Fig.~\ref{fig:dsf}(c,d) for $\phi/\pi=0.01$ and Fig.~\ref{fig:dsf_app} for $\phi/\pi=0.05$,
we observe an equally wide continuum suggesting a likewise description of the high-energy excitations in terms of Majorana fermions.
However, neither P nor NP is close to the flux-free state of the KSL and,
in fact, fluxes are no longer conserved quantities in the presence of a magnetic field or $\Gamma$.
Consequently, $\mathcal S(\mathbf{k},\omega)$ at lower energies is distinct from that of the KSL.
More precisely, the continuum does not have a constant spectral gap, dispersion evolves,
and within P a distinctive magnon band appears.

Upon further increasing $\phi$ and thus $\Gamma/|K|$,
$\mathcal S(\mathbf{k},\omega)$ exhibits enhanced dispersions at both the lower and the upper edges
of the continuum in NP and P.
Among the values of $\phi$ examined here, the magnon bands at $\phi/\pi=0.25$ has the largest bandwidth in the P phase.
In the pure $\Gamma$ model (i.e.~$\phi/\pi = 0.5$) the dispersion is still present but its bandwidth is reduced.

Within NP at $\phi/\pi = 0.05, 0.15$, the excitation gap decreases at the $M$ point together with a significant redistribution of spectral weight there indicating the enhancement of ZZ correlations.
Subsequently, a small perturbation like a negative $\Gamma'$ may stabilize a long-range ZZ ordered phase.
The LSWT spectrum at $\phi/\pi = 0.05$ exhibits the same phenomenology due to the nearby transition into the ZZ phase present in the classical model.

Along the lines of accessible momenta not presented here,
we observe that the gap between the two magnon bands closes at the $K$-point
at $\phi/\pi = 0.15$, both in LSWT and tMPO.
This is due to a duality transformation that exists in the parameter space of the $J K \Gamma \Gamma'$ in a $[111]$ field
mapping the $K \Gamma$ model at $\phi = \tan^{-1} (1/2) \approx 0.148\pi$
to the pure FM Heisenberg model at high field \cite{mcclarty_topological_2018}.
For smaller as well as larger $\phi$, the magnon bands are well separated and known to be topological with non-zero Chern numbers
implying edge modes on geometries with open boundaries \cite{mcclarty_topological_2018,joshi_topological_2018}.
Moreover, the magnetic moment just above the critical field
is expected to be fully polarized only at the $\phi$
dual to FM Heisenberg, where the fully polarized stated is in eigenstate.
Anywhere else, frustration leads to a reduced magnetic moment which approaches
the fully polarized state in the limit $h \longrightarrow \infty$.
This motivates to distinguish the longitudinally-polarized paramagnet
in the quantum model from the fully polarized phase in the classical model.

At $\phi/\pi=0.5$, i.e.~the pure $\Gamma>0$ model, the P phase follows the same phenomenology as at smaller $\phi$, see Fig.~\ref{fig:dsf}(e,f) and Fig.~\ref{fig:dsf_app}.
The dispersing magnon bands shift downwards in energy and the excitation gap closes at the $\Gamma$ and $\Gamma'$ point upon the transition into the NP phase.
As the field is lowered, a multi-magnon continuum starts to form within P and persists across the transition into NP.
In contrast, the pure Kitaev model in a $[111]$ field exhibits a band that simultaneously flattens and approaches zero in the entire reciprocal space \cite{gohlke_dynamical_2018}.
At $\phi/\pi=0.5$ and $h \in [1.4, 1.8]$, the classical ground state is some 6-site magnetic order distinct from P or FF, which is why the LSWT spectra are not plotted.

\section{Additional Plots of Linear Spin Wave Dispersion} \label{app:additionallswt}
\begin{figure*}
\subfloat[]{\label{phi010h0778}
\includegraphics[scale=0.3]{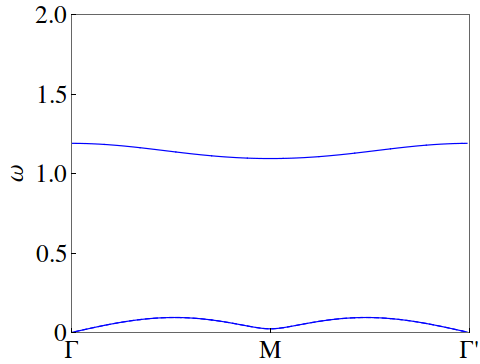}} \quad
\subfloat[]{\label{phi010h080}
\includegraphics[scale=0.3]{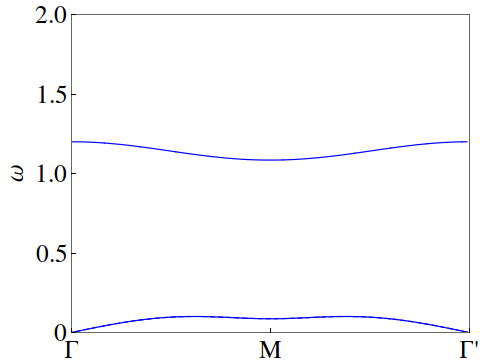}} \quad
\subfloat[]{\label{phi010h084}
\includegraphics[scale=0.3]{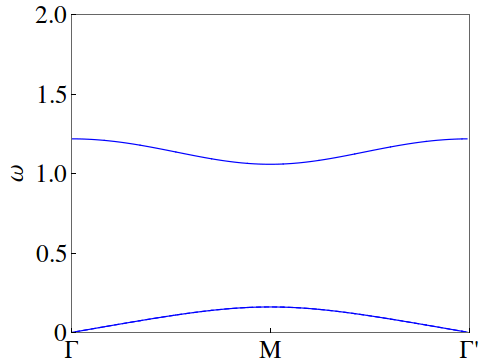}}
\caption{The magnon dispersions in the FF phase calculated from the linear spin wave theory for the $K \Gamma$ model at $\phi/\pi=0.1$ and (a) $h=0.389$, (b) $h=0.4$ and (c) $h=0.42$. Upon lowering the field, the excitation gap at the M point decreases and approaches zero, as there is a nearby phase transition to the ZZ order in the classical model.}
\end{figure*}
The classical limit of the $K \Gamma$ model exhibits a phase transition from the FF phase to the ZZ long-range order upon lowering the field for a wide range of $\phi$ (see Ref.~[\onlinecite{chern_magnetic_2020}] for details).
We examine the dispersion obtained via LSWT within the FF phase near this transition.
We find that, as the field decreases, the excitation gap at the M point shrinks and approaches zero,
as shown in Figs.~\ref{phi010h0778}-\ref{phi010h084}. This is also seen in the DSF of the quantum model, as discussed in Sec.~\ref{scn:dyn} and Appendix \ref{app:additionaldsf}.
We remark that the DSF at a certain choice of $h$ agrees better with the LSWT spectrum at a lower $h$ than that at the same $h$.

%%%%%%%%%%%%%%%%%%%%%%%% Bibliography #################################################

\bibliography{KGGp_field_paper.bib}

\end{document}